# NUMERICAL MODELLING OF WIND-WAVES .
## PROBLEM AND RESULTS

## V. G. Polnikov[1]

**Abstract**:   Stochastic wind sea is an intermediate small-scale physical process responsible for the state of the atmospheric boundary layer and the water upper layer, having dynamics of all scales. To describe behavior of this system, one could use the mathematical formalization based on a spectral evolution model for wind waves. To this end, it needs a well-designed numerical model derived from the principal physical equations. On this way certain theoretical problems take place. At present some of these problems are solved, that gives possibility to construct a lot of numerical wind wave models, the latest version which was proposed in Polnikov(2005). With the aim of assessing real merits of the new source function proposed in the mentioned paper, the latter was tested and validated by means of modification the well known model WAVEWATCH-III. Assessment was done on the basis of comparing the wave simulation results obtained by both models for a given wind field against the buoy data gotten in the three oceanic regions.

   Estimations of simulation accuracy were obtained for three parameters of wind waves: significant wave height, $H_s$, peak wave period, $T_p$, and mean wave period, $T_m$. Comparison of these estimations between the original and modified model WAVEWATCH was fulfilled and analyzed. Advantage of the modified model was revealed, consisting in an increase of simulation accuracy for $H_s$ in 1.2-1.5 times for more than 70% of buoys considered. Additionally, it was found that the speed of calculation was increased in 15%.

## 1.  INTRODUCTION

 Let us consider a typical scheme of the air-sea interface. In simplified approach it consists of three items (Fig. 1):

- Turbulent air boundary layer with the shear mean wind flow having a velocity value $W_{10}(x)$ at the fixed horizon z =10m;
- Wavy water surface;
- Thing water upper layer where the turbulent motions and mean shear currents are present.

The main source of all mechanical motions of different space-time scales at the air-sea interface is a mean wind flow above the surface, which has variability scales of the order of thousand meters and thousand seconds. The turbulent part of a near-water layer (boundary layer) has scales smaller than a meter and a second.  Variability of the wavy surface has scales of tens meters and ten seconds, whilst the upper water motions have a wide range of scales covering all mentioned values. Thus, the wind impacts on the water upper layer indirectly via the middle scale motions of wind waves, and this impact is spread through a wide range of scales.

[1] Research professor, A.M. Obukhov Institute for Physics of Atmosphere of the Russian Academy of Sciences, Moscow, Russia 119017, e-mail: polnikov@mail.ru

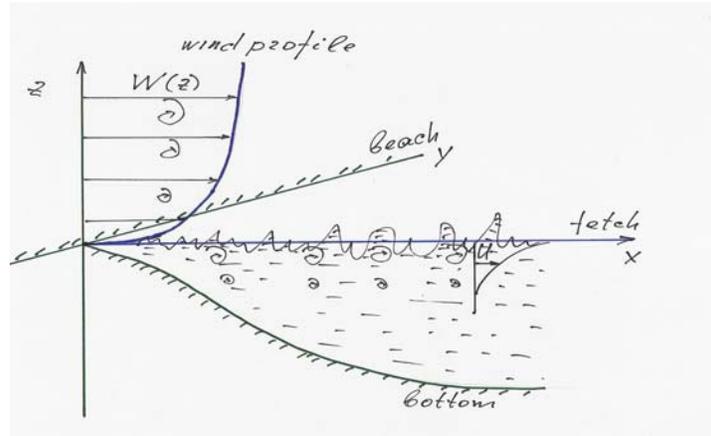

Fig. 1.  The air-sea interface system

Besides of the said importance of wind waves, this phenomenon has its own scientific and practical interest. The former is provided by a physical complexity of the system, whilst the latter is due to dangerous feature of the phenomenon. All the said justify the long period interest to the problem of wind wave modeling, staring from the well know paper by Stokes (1847).

From scientific point of view it is important to describe in a clear mathematical form a whole system of mechanical interactions between items mentioned above, responsible for the exchange processes at the air-sea interface. This is the main aim of the interface hydrodynamics. From practical point of view a mathematical description of these processes permits to solve a lot of certain problems. As an example of such problems one may point out an improvement of wave and wind forecasting, calculation of heat and gas exchange between atmosphere and ocean, surface pollution mixing and diffusion, and so on.

Direct mathematical description of mechanical exchange processes in the system considered is very complicated due to multi-scale and stochastic nature of them (for example, see Kitaigorodskii & Lamly, 1983). It can not be done in an exact form. Nevertheless, real advantage in this point can be reached by consideration of the problem in a spectral representation. Up to the date a principal physical understanding exchange processes at the air-sea interface was achieved to some extent (Proceedings of the symposium on the wind driven air-sea interface, 1994; 1999), and mathematical tool for their description in spectral representation was constructed (for example, see Phillips, 1977). Thus, one may try to make description of main processes from the united point of view, as far as the spectral wind wave model is constructed.

Let us see the main theoretical procedures needed to manage this problem.

## 2.  SPECIFICATION OF THE PROBLEM

All modern numerical models for wind waves are based on the solution of evolution equation for a two-dimensional wave energy spectrum, $S(\sigma, \theta, \mathbf{x}, t)$, (or wave action one, $N(...) = S(...) / \sigma$) given in the space of wave frequency, $\sigma$, and wave propagation angles, $\theta$, which is spread through the geographic coordinates, x, and time, $t$. In general, this equation has the kind

$$\frac{dS(\sigma,\theta,\mathbf{x},t)}{dt} = F(S,\mathbf{W},\mathbf{U}) = In + Nl - Dis \tag{1}$$

Here, the left hand side is the full derivative of the spectrum with time, and the right hand side is the so called source function, $F$, depending on both the wave spectrum, $S$, and the external wave-making factors: local wind, W(x,$t$), and local current, U(x,$t$).

The source function is the "heart" of the model. It describes certain physical processes included in the model representation, which determine mechanisms responsible for the wave spectrum evolution (Efimov& Polnikov, 1991; Komen et al, 1994). It is commonly used to distinguish three terms in function $F$: the atmosphere-wave energy exchange mechanism, $In$; the energy conservative mechanism of nonlinear wave-wave interactions, $Nl$; and the wave energy loss mechanism, $Dis$, related mainly to wave breaking and interaction of waves with turbulence of the water upper layer and the bottom. Differences in representation of the source function terms mentioned above determine general differences between wave models. In particular, the models are classified with the category of generations, by means of ranging the parameterization for $Nl$-term (The SWAMP group, 1985). This classification could be extended, taking into account all source function terms (for example, see Polnikov, 2005; Polnikov, Tkalich, 2006).

Differences in representation of the left hand side of evolution equation (1) and in realization of its numerical solution are mainly related to the mathematics of the wave model. Such a kind representation determines specificity of the model as well. But it is mainly related to the category of variation the applicability range of the models (i.e. accounting for a sphericity of the Earth, wave refraction on the bottom or current inhomogeneity, and so on). We will not dwell on this issue.

What are the main problems in construction of the source function for any numerical model? These problems are related to the derivation of exact forms for the source function terms from the principal physical equations. The latter, in the simplest form valid for an ideal liquid, are as follows

$$\rho\frac{d\mathbf{u}}{dt} = -\vec{\nabla}_3 P - \rho\mathbf{g} + \mathbf{f}(\mathbf{x},t) \quad , \tag{2}$$

$$\vec{\nabla}_3(\mathbf{u}) = 0 \quad , \tag{3}$$

$$u_z\Big|_{z=\eta(\mathbf{x},t)} = \frac{\partial\eta}{\partial t} + (\mathbf{u}\vec{\nabla}_2\eta) \quad , \tag{4}$$

$$u_z\Big|_{z=-H(\mathbf{x})} = \left(\mathbf{u}\vec{\nabla}_2 H(\mathbf{x})\right) \quad . \tag{5}$$

Here $\mathbf{u}$ is the vector of currents in the water upper layer, $\eta(x,t)$ is the water surface elevation function, $P(x,t)$ is the pressure, and typical derivatives notions are used. Remind that Eq. 2 is the maid dynamic equation (in the wind wave task it is used at the water surface z = $\eta(x,t)$ ), Eq. 3 is the mass conservation law, Eq. 4 is the kinematical boundary condition, and Eq. 5 is the condition at the bottom (Komen et al, 1994). Thus the problem is to derive all source terms from the set of equations (2-5), taking into account stochastic feature of motions.

It is easy to understand that the posed problem is quite complicated. Nevertheless, it can be solved under some approximations, if one takes into account each evolution mechanism separately. The history of such investigations is described in quite numerous papers, the main results of which are accumulated in numerous books (Komen et al, 1994; Young, 1999; and others). The main conclusions of these results are as follows.

## 2.1. Nonlinear term

The nonlinear term, $Nl$, is theoretically the most investigated. Under some reasonable suggestions, this term is described by the so-called four wave kinetic integral (Hasselmann, 1962)

$$Nl[N(\mathbf{k_4})] = 4\pi \int d\mathbf{k_1} \int d\mathbf{k_2} \int d\mathbf{k_3} M^2(\mathbf{k_1},\mathbf{k_2},\mathbf{k_3},\mathbf{k_4}) \times$$

$$[N(\mathbf{k_1})N(\mathbf{k_2})(N(\mathbf{k_3}) + N(\mathbf{k_4})) - N(\mathbf{k_3})N(\mathbf{k_4})(N(\mathbf{k_1}) + N(\mathbf{k_2}))] \times$$
$$\times \delta(\sigma(k_1) + \sigma(k_2) - \sigma(k_3) - \sigma(k_4))\delta(\mathbf{k_1} + \mathbf{k_2} - \mathbf{k_3} - \mathbf{k_4})$$  (6)

Here $\mathbf{k}_i$ is the wave vector corresponding to a proper frequency-angular wave component $(\sigma_i, \theta_i)$ ($i = 1,2,3,4$), $M(\dots)$ are the matrix elements describing intensity of interactions between four waves, and $\delta(...)$ is the Dirak's delta-function providing the resonance feature of interactions. This integral is very complicated for numerical solution and cannot be used directly in a numerical model. Thus, for practice, it needs to find an optimal approximation of the kinetic integral, which conserves all its properties.

In our recent study (Polnikov and Farina, 2002), among different theoretically well substantiated approximations the most efficient one (in terms of accuracy and speed of calculation) is the discrete interaction approximation (DIA) proposed by Hasselmann et al. (1985). Just this approximation is used in WAM (WAMDIG, 1988) and WAVEWATCH (Tolman, 1991, Tolman and Chalikov, 1996). Herewith, Polnikov and Farina (2002) and Polnikov (2003) have shown an existence of more exact and more efficient configurations that could replace the original DIA configuration proposed in Hasselmann et al. (1985). In addition to this, Polnikov and Farina have proposed a procedure of the so-called fast version of the approximation (FDIA), which saves time of $NL$-term calculations more than in two times. The latter effect is provided by refusing interpolation procedures, used in DIA and related to necessity to secure exactly the interaction resonance conditions (see references). Taking into account the two advantages mentioned above, in our version of model just the FDIA is used for a numerical representation of $Nl$ term.

In the new source function, an optimized version of the Discrete Interaction Approximation (DIA) is used for parameterization of $Nl$, instead of the ordinary DIA used in WW. The optimization includes the following items:

(a)   Fast version of DIA  (Polnikov and Farina, 2002)

(b)   New, more effective configuration of four-wave interacting waves (Polnikov, 2003).

These two improvements lead to an increase of speed of calculation and a better correspondence of approximate calculations of $NL$-term to the "exact" numerical values of the latter (see original papers).

The fast DIA is governed by the following ratios.

(1)   The calculating frequency-angular grid, $\{\sigma_i,\ \theta_j\}$, is defined typically:

$$\sigma(i) = \sigma_0 e^{i-1} \quad (1 \le i \le N), \quad \text{and} \quad \theta(j) = -\pi + (j-1)\cdot\Delta\theta \quad (1 \le j \le M),$$  (7)

where $\sigma_0$, $e$, and $\Delta\theta$ are the grid parameters specified below.

(2) The first, reference wave component, $(\sigma, \ \theta)$, is located at a current node of the grid (2).

(3) Other 3 waves have the components located at nodes of the same grid, defined by the ratios

$$\sigma_1 = \sigma e^{m1}, \qquad \sigma_2 = \sigma e^{m2}, \qquad \sigma_3 = \sigma e^{m3}, \tag{8a}$$

$$\Delta\theta_1 = n1\,\Delta\theta, \ \ \Delta\theta_2 = n2\,\Delta\theta, \quad \Delta\theta_3 = n3\,\Delta\theta, \text{(where } \Delta\theta_i \equiv \left| \theta_i - \theta \right| \text{)} . \tag{8b}$$

Thus, the optimal configuration of four interacting waves is given by the certain set of integers: $m1$, $m2$, $m3$; $n1$, $n2$, $n3$, which, in turn, are to be especially calculated, in dependence on the grid parameters, $e$ and $\Delta\theta$ (Polnikov and Farina, 2002).

For the frequency-angle grid with parameters $e = 1.1$ and $\Delta\theta = \pi/12$, which are typical for the model WW, the most effective configuration is given by the following parameters of configuration (Polnikov, 2003)

$$m1 = 3, \ m2 = 3, \ m3 = 5; \qquad n1 = n2 = 2, \ n3 = 3. \tag{9}$$

Finally, $Nl$-term is calculated by the standard formulas, making the loops for the reference components, $(\sigma, \ \theta)$, arranged through the grid (7) . For completeness, these formulas are as follows

$$Nl(\sigma,\theta) = Nl(\sigma_3,\theta_3) = I(\sigma_1,\theta_1,\sigma_2,\theta_2,\sigma_3,\theta_3,\sigma,\theta\,), \tag{10a}$$

$$Nl(\sigma_1,\theta_1) = Nl(\sigma_2,\theta_2) = -I(\sigma_1,\theta_1,\sigma_2,\theta_2,\sigma_3,\theta_3,\sigma,\theta\,), \tag{10b}$$

where

$$I(...) = C_{nl}\,g^{-4}\sigma^{11}\Big[ S_1 S_2 (S_3 + (\sigma_3/\sigma)^4 S\,) - S_3 S\,((\sigma_2/\sigma)^4 S_1 + (\sigma_1/\sigma)^4 S_2)\Big]. \tag{11}$$

Here, $C_{nl}$ is the only fitting nondimensional coefficient, and notation $S_i \equiv S(\sigma_i,\theta_i)$ is used.

## 2.2. Input term

Theoretical grounds for representation of the input term in spectral form were given in Phillips (1957) and Miles (1957). Since that time a lot of authors have contributed into theoretical solution of the problem, but it is not found in the final form yet. Therefore, the most of recognized parameterizations of the input term are based on the representation of the kind

$$In = \beta\,(\sigma, \ \theta, \mathbf{U})\,\sigma\,S\,(\sigma, \ \theta) \tag{12}$$

which corresponds to the Miles' mechanism of wave generation by the wind field $\mathbf{U}(\mathbf{x},t)$. The Phillips' mechanism is important for very early stage of wave generation and usually is not used in practical models.

In the aspect of the input term description the main problem is to specify the kind of the wave growing increment $\beta\,(\sigma, \theta, \mathbf{U})$ as a function of its arguments. To do this, one usually attracts the reliable empirical data (Snyder et al., 1981; Plant, 1982; and others). Last decade some theoretical results of numerical simulations for the boundary layer are attracted as well (Makin and Chalikov, 1980; Chalikov, 1980; Janssen, 1989, 1991; Chalikov and Belevich, 1993; Makin and Kudryavtzev, 1999; and others). Here it is worth while to mention that despite of numerous theoretical simplifications, some of these results are more informative than experimental ones, though detailed discussion of this question is out of present consideration.

Among numerous points of the problem, only three ones will be discussed here. They are as follows

- a size of a frequency interval where a parameterization of input term is valid;
- existence a frequency domain where the input term is negative;
- a kind of wind representation which should be used for scaling of the input term: a wind at the certain fixed horizon $h = z$ above the mean water level , U(z), or a friction wind velocity, $u_*$, given by the ratio

$$u_* = C_d^{1/2}(z)U(z) \tag{13}$$

where $C_d(z)$ is the drag coefficient for the horizon $z$.

All other points, namely, angular dependence, dependence of $u_*$ on wave age $A$, and other related questions will not be discussed here for the reason of high extent of their uncertainty.

The widely used parameterization of the wave growing increment proposed in Snyder et al. (1981). It looks like

$$\beta(\sigma, \theta, \mathbf{U}) = \max\left[0, a\ \frac{\rho_a}{\rho_b}\left(\frac{U_{10}\sigma}{g}\cos(\theta - \theta_u) - b\right)\right]. \tag{14}$$

Here the following notions are used: $\rho_a$ and $\rho_b$ is the air and water density, respectively, $g$ is the gravity acceleration, $a$ and $b$ are the fitting parameters, and $\theta_u$ is the local wind direction. Parameters $a$ and $b$ are varying in the following intervals: $a \cong 0.2$-$0.3$ и $b \cong 0.9$-$1$.

The value of the latter parameter, $b$, depends on the representation of the wind. Significance of a value of $b$ is provided by the fact that, according to representation (14), it defines the lower cutting frequency of the input term, $\sigma_L \cong bg/U_{10}$. Below this low limit frequency increment $\beta$ is equal to zero. But, as it was shown in Chalikov (1980), Makin and Chalikov (1980), Chalikov and Belevich (1993), the latter feature of representation (10) is not correct from physical point of view. In these papers it was shown that theoretically expected result is a small negative value of $\beta$ in the domain $\sigma \leq g/U(10)$.

The other restriction of representation (10) is that the upper limit of its validity (estimated still in Snyder et al. (1981)) is given by the ratio

$$\frac{U_s\sigma}{g} \leq 3. \tag{15}$$

Though for many practical aims this frequency limit is sufficient, in theoretical papers (for example, Chalikov and Belevich, 1993; Makin and Kudryavtzev, 1999) it was shown that for better description of boundary layer evolution, a more wider frequency interval is important. Thus, representation (14) should be changed by another one with a wider interval of validity.[2]

---

[2] Note that the form (14) of the input term is used till present (with some corrections it is used in WAM (Komen et al., 1994)).

The latter theoretical request could be met by using, in addition to representation (14), the following approximation (Plant, 1982)

$$\beta = (0.04 \pm 0.02) \left( \frac{u_* \sigma}{g} \right)^2 \cos(\theta - \theta_w) \qquad (16)$$

where $u_*$ is the friction velocity, and $\theta_w$ is the local wind direction.

In terms of $U(10)$, representation (16) is valid in the interval[3]

$$2.5 \le U(10) \sigma / g \le 75 . \qquad (17)$$

So, the better way is to combine representations (14) and (16). Just this work was done in Yan (1987) where the following combined representation was proposed:

$$\beta = \left\{ \left[ 0.04 \left( \frac{u_* \sigma}{g} \right)^2 + 0.00544 \frac{u_* \sigma}{g} + 0.000055 \right] \cos(\theta - \theta_w) - 0.00031 \right\} . \qquad (18)$$

In the present optimized version, parameterization of $\beta$ has the kind (Polnikov, 2005)

$$\beta = C_{in} \max \left\{ -b_L, \left[ 0.04 \left( \frac{u_* \sigma}{g} \right)^2 + 0.00544 \frac{u_* \sigma}{g} + 0.000055 \right] \cos(\theta - \theta_w) - 0.00031 \right\} \qquad (19)$$

Specific feature of this approximation is an existence of negative value of the increment, $\beta = -b_L$, corresponding to the waves propagating with the velocity greater than the properly directed projection of the wind velocity. This feature of $\beta$ is physically important, what was proved in numerical studies by Chalikov (for reference, see Tolman and Chalikov, 1996). Coefficient $C_{in}$ and parameter $b_L$ are the subjects of the model fitting. Default value for the latter is $b_L = 5 \cdot 10^{-6}$.

In this work, a transition $W_{10} \Leftrightarrow u_*$ is done by the methodic incorporated into the WW's codes. But, in principle, it could be calculated by the special dynamic boundary layer block, which may be included into the model (see Polnikov, 2005). This is the task for a further elaboration of the model.

## 2.3. Dissipation term

The dissipation term is the least investigated. Empirical observations of energy dissipation processes in wind waves are rather numerous (for example, see Banner and Tian, 1998; Donelan, 1998; Babanin et al, 2001) but formal mathematical description of them in a spectral form is not well understood. For these reasons there is not a widely recognized parameterization for the dissipation term.

In WAM the following quasi-linear parameterization of $Dis$ is used

$$Dis = \gamma(\sigma, \theta, \mathbf{U}, E) \sigma S(\sigma, \theta), \qquad (20)$$

---

[3] The value $C_d(10) = 0.002$ is taken for this estimation.

the principal form of which was founded in Hasselmann (1974) (the so-called "whitecapping" mechanism). Here $E$ is the total energy of waves. We will not dwell on this obsolete parameterization, the shortages of which were discussed in Banner and Young (1994). We only mention that the latter authors proposed some modifications of the form (20) which improved WAM radically (see also Alves, 2000; Makin & Stam 2003). But these modifications of (20) are rather formal and hardly could be founded theoretically.

In this aspect more substantiated approach was proposed in Tolman and Chalikov (1996). They shared the frequency interval into two parts: low frequency domain, $\sigma \leq 2.5\sigma_p$, and high frequency one, $\sigma > 2.5\sigma_p$ (where $\sigma_p$ is the peak frequency), supposing that a dissipation mechanism is different in a different domain. In the high frequency domain they put "whitecapping" mechanism of the Hasselmann's type like (20) with a strong dependence of $\gamma$ $(\sigma,\theta,\mathbf{U},E,A)$ on the energy level of the equilibrium spectrum intensity, $A$. In the low frequency domain they proposed the "eddy viscosity" mechanism of the kind

$$Dis = -\left[u_* h(E)\phi(A)\right]k^2 S(\sigma,\theta)$$ (21)

where expression in the square brackets is an analog of the upper layer viscosity (for details see original paper). Just this theoretical idea is very fruitful. At present the Tolman-Chalikov parameterization (TCH) of $Dis$ (with some sophistication) is used in WAVEWATCH.

The main theoretical problem of the TCH approach is how to estimate the "eddy viscosity". In addition to this they have a technical problem: how to joint two mechanisms of dissipation? In the original paper these problems were solved formally (in a rather arbitrary way without any mathematical and physical grounds). Just this is the main shortage of the TCH parameterization of the dissipation term.

Herewith, the idea of using the viscosity mechanism of wind wave dissipation was proposed in Efimov and Polnikov (1986) many years ago. It was used in Polnikov (1991) for construction of alternative model and theoretically substantiated in Polnikov (1994). The essence of this theory is the following.

We start from the well known formula for viscosity dissipation

$$Dis(\sigma,\theta,S,\mathbf{U}) = \nu_T(\sigma,\theta,S,\mathbf{U})k^2 S(\sigma,\theta)$$ (22)

which is applied to the energy spectrum of waves, $S(\sigma,\theta)$. Here the function $\nu_T$ has a meaning of the spectral representation of viscosity for the upper layer of waving water. In our case, it is assumed to be due to a small-scale turbulence of the upper layer, which, in turn, is provided by different kinds of dissipation processes (breaking of waves, whitecapping, sprinkling, shear flows, etc). From theoretical point of view our aim is to express these small-scale processes (i.e. $\nu_T$) via the wave-scale process parameters.

To this end, we assume (with some simplification) that function $\nu_T$ can be expressed in terms of the wind-wave system parameters $\sigma$, $\theta$, g, $\mathbf{U}$, and $S(\sigma,\theta)$, only. Taking into account the fact of presence a small non-dimensional parameter of the system, $\alpha = S(\sigma,\theta)\sigma^5/g^2 \cong 10^{-2}$, in our case, the most general form of $\nu_T(\sigma,g,\mathbf{U},S)$ may be written in the kind

$$\nu_T = C(\sigma,\theta,g,\mathbf{U})\sum_{n=0}^{N}\nu_n(\sigma,\theta,g,\mathbf{U})\alpha^n(\sigma,\theta).$$ (23)

Herewith, the series (23) could be restricted by a few first terms ($N$ = 2-3) without any lost of generality (for details, see Polnikov, 1994). The choice of a certain form for the fitting function $C(\sigma, \theta, g, \mathbf{U})$ could be done from the fact of existence of a stable (equilibrium) shape for the fully developed wind-wave spectrum, $S_{eq}(\sigma, \theta)$, in the high frequency domain, $\sigma > 2.5\sigma_p$. In this domain the spectrum shape should obey to the following spectrum stabilization condition

$$F\Big|_{S=S_{eq}} = \big[Nl + In - Dis\big]\Big|_{S=S_{eq}} \approx 0 \,. \tag{24a}$$

As far as functions $Nl\ (S)$ and $In\ (\sigma, g, \mathbf{U}, S\ )$ are already known (see above), specification of $C(\sigma, \theta, g, \mathbf{U})$ could be done in an explicit kind, if we propose the following simplifications.

Firstly, we restrict our representation of the total source function $F$ by a function of the third order in wave spectrum $S$. A linear term of this function can be ascribed to function $In(S)$, whilst the third order one is ascribed to function $Nl(S)$. Thus, the most important term of function $Dis(S)$ is one of the second power in spectrum $S$.

Note that linear and cubic summands of $Dis(S)$, provided by formulas (22), (23), in principle, cannot be separated from the proper terms in $In(S)$ and $Nl(S)$ for a formal numerical representation of source function $F$, as far as for each term we use fitting coefficients accumulating all summands of the same power in $S$. For these reason it is not worth while to seek for them.

Secondly, to simplify determination of $Dis(S)$, we exclude the term $Nl$ from the condition (24a). It is due to the fact that $Nl$ contribution to function $F$, in high frequency domain ($\sigma > 2.5\sigma_p$), is not more than 10-15% with respect to the other summands of the source function (see, for example, estimations in Komen et al., 1994; Efimov and Polnikov, 1991). So, we can replace equation (27a) by the following one

$$\big[In - Dis\big]\Big|_{S=S_{eq}} \approx 0 \,. \tag{24b}$$

Thirdly, for the determination of the small scale parameter of the system $\nu_T(\sigma, \theta, g, \mathbf{U}, S)$ one could attract semi-phenomenological theory. Two versions of this theory could be found in Polnikov (1994, 1995).

Thus the theoretically substantiated expression for dissipation term in deep water has the kind

$$Dis(\sigma, \theta, S, \mathbf{U}) = \widetilde{\gamma}(\sigma, \theta, Ex)\frac{\sigma^4}{g^2}S^2(\sigma, \theta) \,. \tag{25}$$

The principal feature of representation (25) is the second power of $Dis(S)$ in the wave spectrum, $S(\sigma, \theta)$. It permits to specify function $\widetilde{\gamma}(\sigma, \theta, Ex)$ by using the condition (24b), if the shape of equilibrium spectrum for wind waves, $S_{eq}(\sigma, \theta)$, is known. But the choice of the latter is not unequivocal, because of real physical uncertainty in the falling law for the spectrum tail at the high frequencies (Rodriguez and Soares, 1999). To make a choice, we take into account that the total transfer of the wind momentum to waves, given by the formula

$$M = \int_{\sigma} d\sigma \oint_{\theta} d\theta \big[(k/\sigma)In(\sigma, \theta)cos(\theta)\big] \sim \int_{\sigma} d\sigma \oint_{\theta} d\theta \big[\sigma^4 S_{eq}(\sigma, \theta)\big] \quad , \tag{26}$$

should be limited. In such case, the preferable choice of the falling law should be of the kind

$$S_{eq}(\sigma,\ \theta) \sim \sigma^{-5}\ ,$$ (27)

which corresponds to the traditional representation of equilibrium spectral form (Phillips, 1957; Komen et al., 1994). In such a case, from (22), (24b), (25) and (27) it follows

$$Dis(\sigma,\theta,S,\mathbf{U}) = c(\sigma,\theta,\sigma_p)\max\left[\beta_L, \beta(\sigma,\theta,\mathbf{U})\right]\frac{\sigma^6}{g^2}S^2(\sigma,\theta)$$ (28)

where a small limiting value of dissipation $\beta_L$ is introduced from physical point of view: dissipation cannot be equal to zero nowhen.

In (28) $\beta(\sigma,\theta,\mathbf{U})$ is given by formula (19), and the non-dimensional fitting function $c(\sigma,\theta,\sigma_p)$ describes fine details of dissipation rate in the vicinity of the peak frequency $\sigma_p$ and specific angular dependence of $Dis$. In present version of the proposed model, the latter function is given by

$$c(\sigma,\theta,\sigma_p) = C_{dis}\max\left[0,\ (\sigma - c_\sigma\sigma_p)/\sigma\right]T(\sigma,\theta,\sigma_p)$$ (29)

and

$$T(\sigma,\theta,\sigma_p) = \left\{1 + 4\frac{\sigma}{\sigma_p}\sin^2(\frac{\theta - \theta_w}{2})\right\}\max\left[1,\ 1 - \cos(\theta - \theta_w)\right].$$ (30)

In the angular function $T$, the first factor is introduced to describe angular dependence of $Dis$ at the high frequencies, mainly, and the second one does the angular dependence for opposing wind. Here we should note that though parameterizations (29), (30) are rather phenomenological, but the main kind of $Dis(S)$ (28) is theoretically well substantiated, and the whole our approach could be classified as semi-phenomenological. Finally, formulas (31), (32), and (33) accomplish our specification of the dissipation term.

In present calculations we use representation of $Dis$ in which $\beta_L$, $c_\sigma$, and $C_{dis}$ are the fitting parameters.

### 2.4. The task of wave models comparison

Three models of the third generation, WAM (The WAMDI group, 1988), WAVEWATCH (Tolman and Chalikov, 1996), and SWAN (Boij et al, 1999), are the most widely spread in the world at present. The first two are mainly used to solve global tasks of the wave forecast in deep water. The third one represents by itself an elaboration of the first model for the case of finite depth water. Mainly, it is used to solve regional tasks.

 The models mentioned are rather well fitted against observations and give satisfactory results. But, they have been constructed on the physical grounds which are more than 10 years old. Therefore, despite of permanent updating, they are obsolete at some extent, both in the aspect of substantiation the source function terms, and in the aspect of technical realization mathematics of the models. All these circumstances restrict potential possibilities of the models. Herewith, a regular appearing new theoretical results and permanent extension domain of the models application dictates necessity of construction a new, more modern model. First of all, it is related to modification of the source function, $F$. One of such a kind modification was proposed in the recent paper (Polnikov, 2005), where it was called as "the optimized source function" (for explications, see original paper). Our attempt to incorporate this, new source function into the model WAM gave very encouraging results (Polnikov at al, 2008). We have found that the errors of numerical simulations were decreased in 1.5-2 times, whilst the speed of calculation was enhanced on 25%.

In present paper, we pose the task to estimate real merits of the new source function by means of incorporating it into the mathematical codes of the model WAVEWATCH-III (version 2.22) (hereafter is referred as WW), as the most advanced one at present (Tolman et al., 2002). This estimation will be done on the basis of comparison the numerical simulations against the buoy measurements of wind waves, gotten in two parts of the World Ocean: Eastern and Western parts of the North Atlantic.

Hereafter, the numerical model WW, modified with replacement of the original source function by the new one, is referred as the model NEW.

## 3. METHODIC OF STUDYING THE MODELS PERFORMANCE

There are two approaches to study the numerical model performance: testing and validation processes. The former is based on execution of academic testing tasks, and the latter does on validation of models against natural observation data. In our study, we dealt with both of the approaches. As far as the basic principles of these processes have their own specifications, it is worth while to remind them briefly, following to Efimov& Polnikov (1991), Komen et al. (1994).

### 3.1. Initial regulations for testing the models

There are three principal features providing for importance of the testing process. They are as follows:

I. Possibility to reveal numerical features of the model by means of simplified consideration provided by using the fully controlled wind and boundary conditions.

II. Message comprehensibility and predictability of the testing tasks.

III. Simple and narrow aimed posing the testing tasks.

There is a long list of testing tasks which could be used for a models properties evaluation (for example, see The SWAMP group, 1985; Efimov and Polnikov, 1991; Komen et al., 1994; or Polnikov, 2005). But execution of all of them is out of our main aim. At present stage of studying, we have used the following list of tests.

#1. Straight fetch test (wave development or tuning test).

#2. Swell decay test (dissipation test).

In general, it is possible to distinguish three levels of adequacy of numerical wind wave models, which are defined by the proper choice of reference parameters used for comparison against observations (Efimov and Polnikov, 1991). But here we restrict ourselves by the first level only, as far as checking of the second and third level of adequacy needs much more time and efforts. It is postponed for the future studies. Example of such a kind testing can be found in Polnikov (2005).

Due to principal role of the test #1, the first level reference parameters are of the most importance too. They are as follows:

a) non-dimensional energy, $\widetilde{E} = \dfrac{Eg^2}{W_{10}^4}$ (or $E^* = \dfrac{Eg^2}{u_*^4}$ ), (31)

b) non-dimensional peak frequency, $\widetilde{\sigma}_p = \dfrac{\sigma_p W_{10}}{g}$ (or $\sigma_p^* = \dfrac{\sigma_p u_*}{g}$ ), (32)

where dimensional energy, $E$, is calculated by the ordinary formula, $E = \iint S(\sigma, \theta) d\sigma d\theta$ , and $\sigma_p$ is the peak frequency of the spectrum $S(\sigma, \theta)$ .

Both values, $\widetilde{E}$ and $\widetilde{\sigma}_p$ , estimated from simulations for the stationary stage of the wind wave field, are considered as functions of the non-dimensional fetch, $\widetilde{X} = Xg / W_{10}^2$ . Numerical dependences $\widetilde{E}(\widetilde{X})$ and $\widetilde{\sigma}_p(\widetilde{X})$, found in simulations, are to be compared with the reference empirical ratios of the kind (Komen et al, 1994):

(a) For the stable atmospheric stratification:

$$\widetilde{E}(\widetilde{X}) = 9.3 \cdot 10^{-7} \widetilde{X}^{0.77} ; \qquad \widetilde{\sigma}_p(\widetilde{X}) = 12 \widetilde{X}^{-0.24} \qquad (33)$$

(b) For the unstable atmospheric stratification:

$$\widetilde{E}(\widetilde{X}) = 5.4 \cdot 10^{-7} \widetilde{X}^{0.94} ; \qquad \widetilde{\sigma}_p(\widetilde{X}) = 14 \widetilde{X}^{-0.28} \qquad (34)$$

For the test #2, proper reference parameters are specified below.

### 3.2. Comparative validation of the models

Another approach of studying the properties of numerical models is the process of validation. But, we deal with atypical validation procedure; it is rather a comparative validation of two models.

In this regard, it is worth while to note that the comparative validation procedure is a delicate methodological process, the main points of which are not well formulated till now. The proper formulations should be formalized as a series of special regulations, which is planned to be done further in a separate work. At present, as the primary initial regulations, we can state that the execution of comparative validation procedure requires meeting to several certain conditions. The main of them are consisting in availability the following items:

a) Reasonable data base, including accurate and frequent wave observations;

b) Reliable wind field, given on a rather thick space-time grid for the whole period of wave observations;

c) Properly elaborated mathematical part of numerical model of the kind (1);

d) Certain numerical wind wave model, chosen for comparison as a reference one.

In our work, the last two requirements were satisfied by the choice of the model WW, whilst the other conditions were met by the following way.

A. Two oceanic areas were chosen, for which the wave observation data were used: Western and Eastern parts of the North Atlantic (hereafter referred as NA).

At the first stage of validation process, we have used the one-month data (January, 2006) for 19 buoys located both in the Western and Eastern parts of NA. These data have a time discrete of 1 hour what corresponds to more than 700 points of observations on each buoy.

B. As the wind field, we have used a reanalysis (made in NCEP/NCAR) with a spatial resolution of $1.0^0$ both in longitude and in latitude. The time resolution for the wind was 3 hours. To exclude uncertainties with the boundary conditions, the simulation region was restricted by the following coordinates: $78^0$S – $78^0$N in latitudes and $100^0$W – $20^0$E in longitudes, and the ice covering fields were included into consideration.

On the basis of these external data, the first stage of validation has been executed. These calculations resulted in a sophisticated choice of the fitting coefficients, $C_{in}$, $C_{dis}$, $C_{nl}$, found for the default values of other fitting parameters mentioned above, $b_L$, $\beta_{dis}$, $c_\sigma$.

At the second stage of validation, we have used the long-period historical data of the National Buoy Data Centrum (NBDC) (covering October-May period of 2005-2006 years) for 12 buoys located in the Western part of NA. The wind fields and the time-space resolution were the same as at the first stage. Basing on these data, the standard validation of the both models has been done, without changing any coefficients.

### 3.3. Specification of numerical simulations

In our calculation we have used the frequency-angle grid of the form (7), having parameters

$$\sigma_0 = 2\pi \cdot 0.04 \text{ rad}, \quad e = 1.1 \quad \text{and} \quad \Delta\theta = \pi/12 \ (\text{or } \Delta\theta = 15^{\text{o}}) \tag{35}$$

with the number of frequency bins of $N =24$ and number of angle bins of $M =24$.

In the case of model testing, the spatial grid was taken in Cartesian coordinates, including 100 points in x-direction and 21 points in y-direction. In the case of model validation in oceanic regions, the grid was taken in spherical coordinates, including the number of points depending on the region (see below). The space and time steps of calculations, $\Delta X, \Delta Y, \Delta t$, were varying in accordance with the tasks. Every time, an initial spectrum was taken in the frame of WW codes.

### 3.4. Statistical measures of the validation errors

To assess an accuracy of simulations for a time-series of a certain wave parameter, *P(t)*, we have used the following error estimates: a root-mean-square error, $\delta P$, given by the formula

$$\delta P = \left( \frac{1}{N_{obs}} \sum_{n=1}^{N_{obs}} \left( P_{num}(n) - P_{obs}(n) \right)^2 \right)^{1/2}, \tag{36}$$

and a relative root-mean-square error, $\rho P$, defined as

$$\rho P = \left( \frac{1}{N_{obs}} \sum_{n=1}^{N_{obs}} \left( \frac{P_{num}(n) - P_{obs}(n)}{P_{obs}(n)} \right)^2 \right)^{1/2}. \tag{37}$$

Here $N_{obs}$ is the total number of observation points taken into consideration, and the evident sub-indexes are used.

In addition to this, the following arithmetic errors are very useful for analysis:

$$\alpha P = \left( \frac{1}{N_{obs}} \sum_{n=1}^{N_{obs}} \left( P_{num}(n) - P_{obs}(n) \right) \right). \tag{38}$$

Here we remind that the first two errors describe statistical scattering of the simulations results (or the errors of input parameters, like a wind), whilst the latter one does the mean shift of numerical results with respect to observations.

There are several other statistical characteristics which could be useful for assessment of the numerical model quality (correlation coefficient, probability function, and so on, for example, see Tolman et al, 2002). But at present stage of validation they are omitted, for the sake of more clearness of the primary analysis of the results presented below.

## 4. RESULTS OF THE MODELS TESTING

### 4.1. Straight fetch test

Pose of the task. Spatially homogeneous and constant in time wind, $W(x, t) = W_{10} = const$, is blowing normally to an infinite straight shore line. Initial conditions are given by a homogeneous wave field with a small intensity of wave spectrum. Boundary conditions are invariable in time and correspond to the initial wave state.

The purpose of the test is to check correspondence of the wind wave growing curves, $\widetilde{E}(\widetilde{X})$, $\widetilde{\sigma}_p(\widetilde{X})$, provided by the model, to the reference empirical growing curves, for the stationary state of developed wind waves, given by ratios (33), (34).

As far as the results of this test are typical and well predicted, here we show only some examples of testing results of the model NEW for different wind values, $W_{10}$ =10-30 m/s. They are presented in Figs. 2, 3, for values of $C_{nl} = 9 \cdot 10^7$, $C_{in} = 0.4$, $C_{dis} = 60$, and the default values for the other fitting parameters. The proper results for original WW are presented, for example, in Tolman and Chalikov (1996).

From Figs 2, 3 one can see that curves $\widetilde{E}(\widetilde{X})$, $\widetilde{\sigma}_p(\widetilde{X})$ provided by the modified model are in a good correspondence with empirical ratios (33). It permits to state a good degree of tuning the model what proves the first level of its adequacy, at least.

Second, it should be taken into account that the empirical dependences (33) are valid for nondimensional fetches of the range $10^2 \leq \widetilde{X} \leq 10^4$, with the errors of the order of 10-15% (Komen et al, 1984). This natural scattering feature of empirical data provides for a possibility to fit a lot of different models to the dependences (33) with the proper accuracy.

Third (and it is of the most importance), a good correspondence of numerical and empirical dependences $\widetilde{E}(\widetilde{X})$, $\widetilde{\sigma}_p(\widetilde{X})$ does not provide for an unequivocal choice of the fitting parameters. Coincidence with the errors of 10-15% can be achieved for a continuum of values for the fitting parameters, like $C_{in}$, $C_{dis}$, $C_{nl}$, and the others mentioned above. This result is provided by the simplified meteorological conditions used in the testing task. The sophisticated fitting of the model could be achieved only by means of the model validation against observations, executed for a rather long period of wave evolution under well controlled, but varying meteorological conditions. This point will be discussed in some details below.

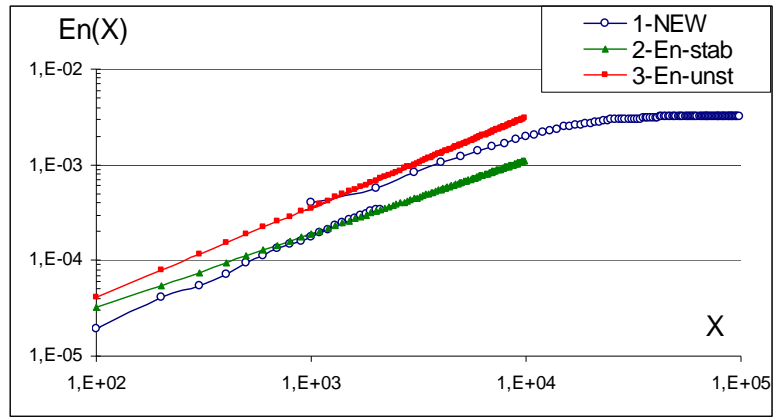

Fig. 2. Dependence of non-dimensional energy on non-dimensional fetch, $\widetilde{E}(\widetilde{X})$, for $W_{10}=$ 10m/s: 1 – model NEW; 2 – Stable stratification ; 3 – Unstable stratification .

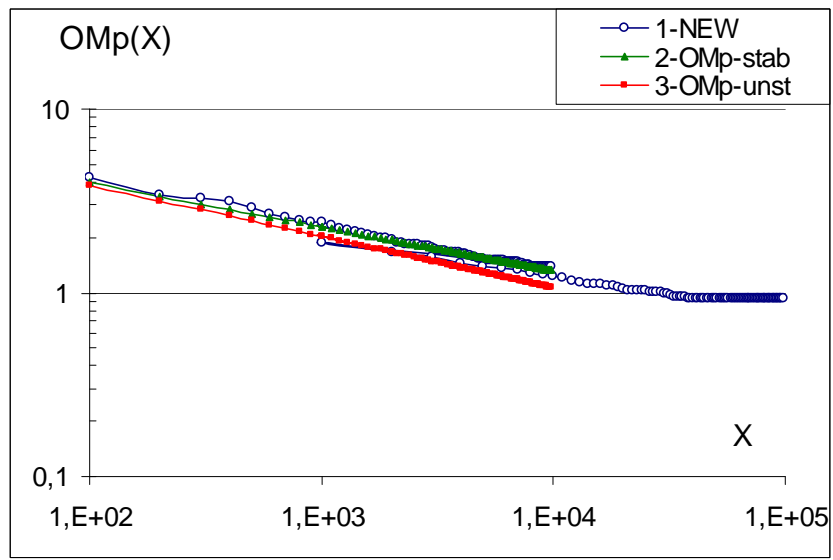

Fig. 3. Dependence of non-dim peak frequency on non-dim fetch, $\widetilde{\sigma}_p(\widetilde{X})$. For legend, see Fig. 3.

## 4.2. Swell decay test

Pose of the task. Forcing wind of the fixed values is present in the first part of the testing area: $W(X) = W_{10}$ at points $0 \leq X \leq X_m$. In the second part of the area, the wind is absent: $W(X) = 0$ at $X_m < X \leq 3X_m$. Initial wave state and boundary conditions are typical (see above the pose of test #1).

The numerical evolution is continued for the period, $T$, providing for a full development of waves at the fetch $X=X_m$, and getting a stable state of the decaying swell field, taking place in the second part of the testing area. Corresponding value of nondimensional time, $\widetilde{T} = Tg / W_{10}$, should be about several units of $10^5$.

Aim of the test is to reveal quantitative features of the swell decay process, starting with different peak frequencies, $f_{sw} = f_p (X_m)$. The latter is considered as a principal initial characteristic of the swell, taking into account that the initial intensity of the swell is mainly provided by $f_{sw}$. To the aim posed, the different values of $W_{10}$, $X_m$, and $T$ should be taken into consideration.

For $W_{10} = 10$ m/s, we took: $\Delta X = 10$ km, $X_m = 240$ km, $T = 48$ h; and for $W_{10} = 20$ m/s we did: $\Delta X = 40$ km, $X_m = 760$ km, $T = 72$ h.

In the second part of the area, the following reference parameters are checked:

- relative energy lost parameter
  $$Ren(X) = E(X\text{-}X_m)/E(X_m); \tag{39}$$
- relative frequency shift parameter
  $$Rf_p(X) = f_p (X\text{-}X_m)/f_p (X_m). \tag{40}$$

As far as there are no widely recognized empirical dependences $Ren(X)$ and $Rf_p(X)$, the found ones are evaluated at the expert level, only. The latter means a quantitative physical analysis (see below).

Results of our simulation are shown in Figs. 4, 5, representing the swell decay process for values $W_{10} = 10$ and 20 m/s. The correspondent values of initial swell frequency, $f_{sw}$, are 0.18 Hz and 0.085 Hz, respectively.

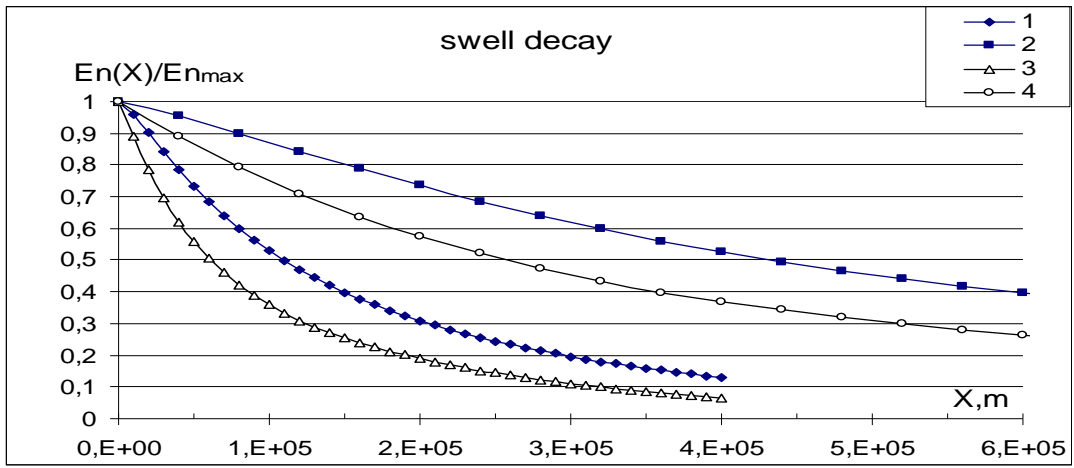

**Fig. 4. Dependence *Ren*(X) for two values of initial peak frequency of swell:**
**1, 2 – original model WW; 3, 4 – model NEW; 1, 3 - $f_{sw}$ =0.18Hz ; 2, 4 - $f_{sw}$ =0.085Hz.**

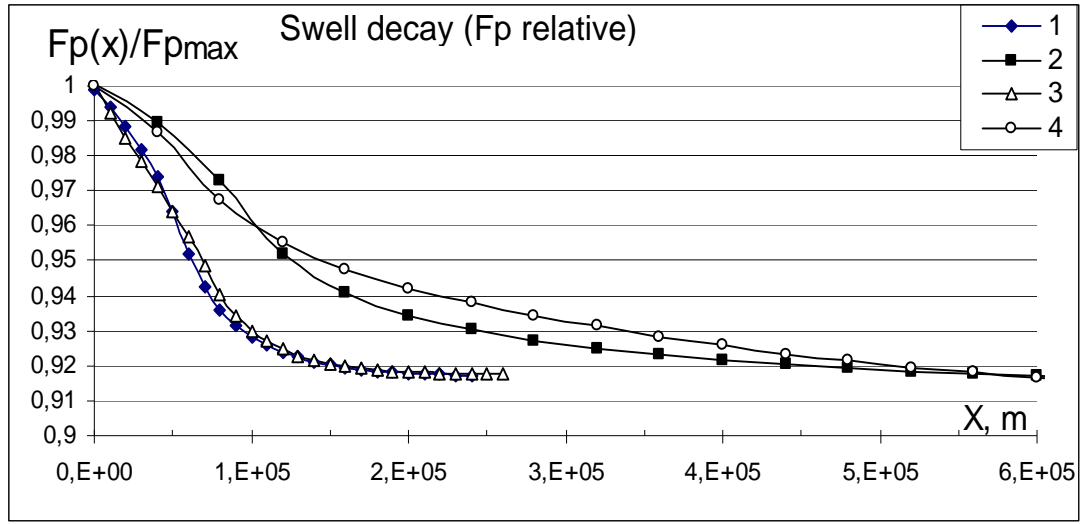

**Fig. 5. Dependence $Rf_p(\mathbf{X})$ for two values of initial peak frequency of swell. F**or the legend, see Fig. 4.

From these figures one can draw the following conclusions.

1) The rate of swell energy dissipation depends strongly on the initial peak frequency of swell, $f_{sw}$. This rate is quickly going down with the distance of swell propagation (Fig. 4).

2) Model NEW has a faster swell dissipation rate (Fig. 4).

3) The rate of peak frequency shifting to lower values, provided by the nonlinear interaction between waves, depends strongly on the initial value of peak frequency, $f_{sw}$ (Fig. 5). The greater $f_{sw}$, the greater rate of frequency shifting. This is well understood, taking into account the formula for nonlinear evolution term.

4) Model NEW has practically the same rate of peak frequency shifting, in contrast to a rate of the relative energy loss (Fig. 5).

This test is very instructive in the physical aspect. Really, from the results obtained, one can draw the following consequences.

First, from the conclusion 2), one can state that the new dissipation term is more intensive than one used in the original model WW.

Second, from the conclusion 4), one can state a very close similarity of the nonlinear terms in the both models.

Third, from previous two consequences, one could state that all qualitative differences of results between these two models are mainly provided by the new parameterization of *Dis*-term. Herewith, we note that though the new parameterization of *In*-terms has a feature of additional background dissipation, in this test, it is two small to play any remarkable role, especially at the initial stage of swell decay.

As one could see later, the last consequence is of the most importance for understanding and treatment of difference between these models, which will be found later during validation.

## 5. RESULTS OF COMPARATIVE VALIDATION OF THE MODELS WW AND NEW

### 5.1. One-month simulations in the North Atlantic

After several runs of the model NEW, intended to a sophisticated choice of the fitting coefficients, $C_{in}$, $C_{dis}$, and $C_{nl}$, we have found that the best results are gained for the following values:

$$C_{nl} = 9 \cdot 10^7, \quad C_{in} = 0.4; \; C_{dis} = 70, \quad \text{and} \quad c_\sigma = 0.7 \tag{23}$$

with the default values of the other fitting parameters.

A typical time history of significant wave height, $H_s(t)$, obtained in these simulations is shown in Fig. 6, for buoy 41001 chosen as an example. From this figure, in particular, one can see that the model NEW does better follow the extreme values of real waves than it is done by the model WW. Visual analysis of all proper curves has showed that this feature of the model NEW is typical for the major part of buoys taken into consideration. More detailed and quantitative analysis needs using the statistical procedures based on the error measures described above.

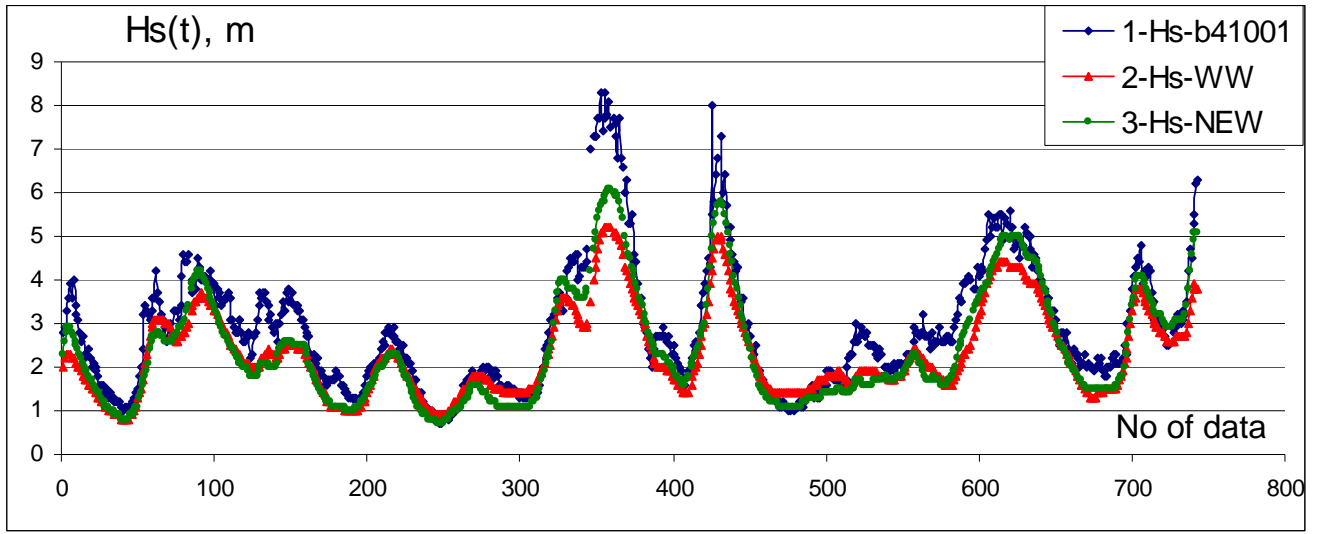

Fig. 6. Time history of the observed and simulated wave heights, $H_s(t)$, on buoy 41001 for
January 2006. 1- wave heights measured on the buoy,
2- wave heights simulated by the model WW, 3- wave heights simulated by the model NEW

At this stage of validation, the properly estimated errors have been found for a significant wave height, $H_s$, only. They are presented in Tabs 1, 2, separately for two parts of NA. For quickness of general (visual) evaluating the results, we have shaded sells corresponding to the cases when the model NEW has a loss of accuracy.

**Table 1.**

**Root-mean-square errors of simulations in the Eastern part of NA**

| Eastern NA, No of buoy | Model WW | | Model NEW | | $\dfrac{(\delta H_s)_{WW}}{(\delta H_s)_{NEW}}$ |
|---|---|---|---|---|---|
| | $\delta H_s$ ,m | $\rho H_s$ ,% | $\delta H_s$ ,m | $\rho H_s$ ,% | |
| 62029 | 0.57 | 14 | 0.54 | 13 | 1.05 |
| 62081 | 0.67 | 15 | 0.56 | 13 | 1.20 |
| 62090 | 0.66 | 14 | 0.57 | 14 | 1.16 |
| 62092 | 0.58 | 14 | 0.53 | 14 | 1.09 |
| 62105 | 0.79 | 18 | 0.68 | 15 | 1.16 |
| 62108 | 0.99 | 15 | 0.84 | 13 | 1.18 |
| 64045 | 0.71 | 12 | 0.61 | 12 | 1.16 |
| 64046 | 0.72 | 15 | 0.76 | 15 | 0.95 |

**Table 2.**

**Root-mean-square errors of simulations in the Western part of NA**

| Western NA No of buoy | Model WW | | Model NEW | | $\dfrac{(\delta H_s)_{WW}}{(\delta H_s)_{NEW}}$ |
|---|---|---|---|---|---|
| | $\delta H_s$ ,m | $\rho H_s$ ,% | $\delta H_s$ ,m | $\rho H_s$ ,% | |
| 41001 | 0.81 | 22 | 0.66 | 20 | 1.23 |
| 41002 | 0.52 | 18 | 0.47 | 18 | 1.11 |
| 44004 | 0.82 | 25 | 0.68 | 26 | 1.21 |
| 44008 | 0.83 | 27 | 0.61 | 24 | 1.36 |
| 44011 | 0.82 | 23 | 0.55 | 18 | 1.49 |
| 44137 | 0.58 | 19 | 0.51 | 17 | 1.14 |
| 44138 | 0.70 | 19 | 0.74 | 19 | 0.95 |
| 44139 | 0.63 | 19 | 0.69 | 20 | 0.91 |
| 44140 | 0.78 | 19 | 0.80 | 19 | 0.97 |
| 44141 | 0.64 | 20 | 0.68 | 20 | 0.94 |
| 44142 | 0.81 | 27 | 0.48 | 18 | 1.69 |

In the aspect of analysis of these results, we should say the following.

First, the accuracy of the model NEW is regularly better with respect to one of the original WW. This result is revealed for more than 70% of buoys considered.

Second, discrepancy of the r.m.s. errors for the both models is remarkable. Typical winning of accuracy for the model NEW is of the order of 15-20%, but sometime it can reach 70% (buoy 44142).

Third, the relative error, $\rho H_s$, calculated by taking into account each point of observations, is not so small (15-27%). It has a tendency of reducing for the model NEW, but it is not so well expressed.

Basing on the said above, we should note that in the present statistical consideration, the relative error $\rho H_s$ is not so sensitive to the specificity of the model, as it could be expected. It seems that the effect of increased sensitivity of $\rho H_s$ could arises, if we introduce the lower limit of wave heights, taken into the procedure of error estimation. For example, the proper error estimations could be done, restricting the time-series points $H_s(t)$ with the wave heights greater than 2 m, only. But, an introduction of limiting values for $H_s$ (or for $T_p$) is not so evident, therefore this point should be especially studied later.

In this connection, it is worth while to mention the accuracy of the input wind. The proper time history for $W_{10}(t)$ is shown in Fig. 7.

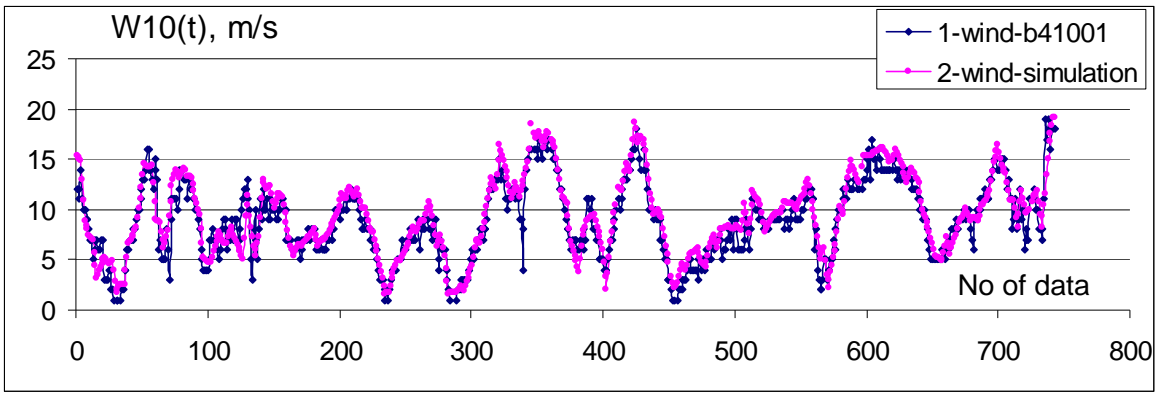

Fig. 7. Time history of the observation and simulation wind , $W_{10}(t)$, on buoy 41001 for January 2006. 1- wind measured on the buoy, 2 – wind used in the modeling simulation

From the first sight, the correspondence between the simulation wind and the observed wind seems to be rather well. But direct calculations of the errors $\delta W_{10}$ and $\rho W_{10}$, made, for example, for buoy 41001, give the values

$$\delta W_{10} = 1.56 \text{ m/s} \qquad \text{and} \qquad \rho W_{10} = 32\% \ . \qquad (41)$$

The first value is more or less reasonable, taking into account that the wind is calculated by reanalysis for the very large domain covering the whole Earth. But the last value, $\rho W_{10}$ in (24), seems to be fairly great with respect to the corresponding relative error $\rho H_s$ (Tab. 1). Due to an arbitrary choice of the buoy considered, one can expect that such a kind mismatch between values of $\rho H_s$ and $\rho W_{10}$ is typical for the present consideration, what, in turn, needs its understanding and explanation.

This mismatch of values for $\rho W_{10}$ and $\rho H_s$ leads to a pose of the following new task: how to treat the present correspondence between these errors. To solve this task, first of all, it needs to have a large statistics of the errors. A part of such a kind statistics will be presented below in Tab. 3. Besides, physically it is reasonable to introduce the lower limiting values for wind, $W_{10}$, and wave heights, $H_s$, which restrict the proper time-series points involved into the procedure of error estimation. In such a way, one could find a physically expected, unequivocal inter-relation between errors $\rho W_{10}$ and $\rho H_s$. If is found, this relation permits to make a proper physical treatment of the errors and to clarify prospective for numerical modeling improvements. Such a kind work is postponed for a future investigation.

### 5.2. Long-period simulations in the Western part of the North Atlantic

Simulating results for the second stage of validation are very similar to the ones presented above. The proper errors are shown the Tab. 3, where the shaded sells correspond to the cases of lost of accuracy in $H_s$ by the model NEW.

From this table, principally, one can state a reasonable advantage of the new model with respect to WW, in the aspect of simulation accuracy for the wave heights, which is defined by the values of r.m.s. error $\delta H_s$. The winning in accuracy is varying in the limits of 1.1-1.5 times.

More detailed analysis results in the following. Arithmetic errors for WW are regularly greater then ones for the model NEW. Herewith, from table 3 it is seen that the model WW gives permanent underestimation of the wave heights, $H_s$, whilst the model NEW has more symmetrical and smaller arithmetic errors. These facts allows us to conclude that the model NEW (and the new source function, consequently) has apparently better physical grounds.

In the aspect of accuracy for the wave periods, $T_m$ and $T_p$, we should confess that the model NEW has less accuracy in calculation of the mean wave period, $T_m$, but, herewith, it has practically the same (or even better) accuracy for the peak wave period, $T_p$ (Tab. 3).

Regarding to the wave periods, we should note a very specific feature, consisting in the fact that the both models show a certain overestimation for the mean wave period, $T_m$, whilst the peak period, $T_p$, is permanently underestimated. The most probable reason of such a behavior of models could be related to an insufficient accuracy for calculation the 2D-shape of wave spectrum, $S(\sigma, \theta)$, taking place for the both models.

As regards to mismatch for the mean wave period, one may additionally suppose that this effect could be related to the methodic of quantitative estimation of $T_m$, realized in the buoy equipment. The systematic error could be provided by the automatic calculation 1D-spectrum of wind waves, $S(\sigma)$, currently (hourly) done with the aim of estimation for $T_m$.

**Consolidated input and output errors for the 8-months simulations in the Western part of NA**

| No of buy/model | $\delta W_{10}$, m/s | $\rho W_{10}$, % | $\delta H_s$, m | $\rho H_s$, % | $\delta T_m$, s | $\rho T_m$, % | $\delta T_p$, s | $\rho T_p$, % | $\alpha W_{10}$, m/s | $\alpha H_s$, m | $\alpha T_m$, s | $\alpha T_p$, s | $\dfrac{(\delta H_s)_{ww}}{(\delta H_s)_{new}}$ | $\dfrac{(\alpha H_s)_{ww}}{(\alpha H_s)_{new}}$ |
|---|---|---|---|---|---|---|---|---|---|---|---|---|---|---|
| 41001/WW | 2.01 | 40 | 0.68 | 22 | 0.93 | 17 | 2.02 | 24 | 0.58 | -0.45 | 0.46 | -1.32 | 1.42 | 2.04 |
| /NEW | | | 0.48 | 18 | 1.23 | 22 | 2.13 | 30 | | -0.22 | 0.79 | -0.88 | | |
| 41002/WW | 1.77 | 48 | 0.48 | 19 | 1.20 | 22 | 2.01 | 27 | 0.25 | -0.23 | 0.78 | -1.05 | 1.09 | 7.67 |
| /NEW | | | 0.44 | 20 | 1.58 | 30 | 2.22 | 35 | | -0.3 | 1.11 | -0.57 | | |
| 41004/WW | 2.54 | 36 | 0.97 | 51 | 1.33 | 31 | 2.40 | 36 | -1.48 | -0.97 | 0.63 | -1.26 | 1.52 | 2.06 |
| /NEW | | | 0.64 | 36 | 1.36 | 32 | 2.38 | 38 | | -0.47 | 0.73 | -1.10 | | |
| 41010/WW | 1.24 | 32 | 0.40 | 19 | 1.61 | 33 | 2.06 | 29 | 0.09 | -0.19 | 1.25 | -0.88 | 1.08 | 2.37 |
| /NEW | | | 0.37 | 20 | 2.07 | 43 | 2.34 | 41 | | -0.08 | 1.69 | -0.21 | | |
| 41025/WW | 2.18 | 50 | 0.44 | 24 | 1.47 | 30 | 2.23 | 30 | 0.47 | -0.05 | 1.09 | -1.16 | 0.81 | 0.29 |
| /NEW | | | 0.54 | 31 | 1.82 | 38 | 2.23 | 35 | | 0.17 | 1.48 | -0.57 | | |
| 41040/WW | 0.91 | 20 | 0.22 | 10 | 1.78 | 30 | 1.87 | 18 | 0.08 | -0.10 | 1.64 | -0.90 | 0.88 | 1.43 |
| /NEW | | | 0.25 | 11 | 2.11 | 35 | 1.96 | 22 | | -0.07 | 1.90 | -0.53 | | |
| 41041/WW | 0.96 | 22 | 0.20 | 09 | 1.92 | 32 | 2.22 | 21 | 0.17 | -0.06 | 1.78 | -0.90 | 0.87 | 1.50 |
| /NEW | | | 0.23 | 10 | 2.26 | 38 | 2.20 | 24 | | 0.04 | 2.06 | -0.54 | | |
| 44004/WW | 1.91 | 40 | 0.72 | 24 | 1.13 | 21 | 1.96 | 24 | 0.16 | -0.38 | 0.52 | -1.34 | 1.26 | 9.5 |
| /NEW | | | 0.57 | 24 | 1.32 | 25 | 1.88 | 26 | | -0.04 | 0.82 | -1.00 | | |
| 44005/WW | 2.28 | 59 | 0.58 | 25 | 1.44 | 30 | 2.27 | 38 | 1.19 | -0.30 | 0.84 | -0.84 | 1.29 | 10.0 |
| /NEW | | | 0.45 | 27 | 1.78 | 37 | 2.24 | 43 | | 0.03 | 1.37 | -0.19 | | |
| 44008/WW | 2.35 | 51 | 0.70 | 25 | 1.11 | 21 | 2.01 | 26 | 0.69 | -0.43 | 0.60 | -1.31 | 1.4 | 4.78 |
| /NEW | | | 0.50 | 21 | 1.30 | 25 | 1.88 | 28 | | -0.09 | 0.91 | -0.90 | | |

Here we should note that in accordance to the definition

$$T_m = \frac{2\pi \int \sigma^{-1} S(\sigma) d\sigma}{\int S(\sigma) d\sigma} \quad , \tag{42}$$

a twice better accuracy of estimation for a spectrum function, $S(\sigma)$, is needed, to meet the proper requirements for accuracy of evaluation for $T_m$. Herewith, it is well known that an accurate estimation of $S(\sigma)$ is not so simple, to be done correctly in a quantitative aspect (Bendat and Piersol, 1971). So, this question needs mode detailed and separate consideration together with the checking documentation of the buoys construction.

Thus, the definite conclusion about superiority of one model against the other can not be drawn at present. Nevertheless, in principle, it could be done later, when the proper criteria will be formulated. This point is only posed here, and we plan to solve it in our future work.

### 5.4. Point of the speed of calculation

By using the numerical procedure PROFILE, we have checked the speed of calculation, realized while execution of all main numerical subroutines used in the models. In terms of consuming-time, the proper time distributions among the main subroutines are shown for the two models in Tab 5. These distributions are corresponding to the case of execution a task of 24-hours simulation of the wave evolution in the whole Atlantic.

**Table 4.**
**Distribution of the central processor consuming-time, realized by the two versions of WW.**

| Model | Name of procedure (*explanation*) | Time, s | Time, % |
|---|---|---|---|
| Original WW | w3snl1md_w3snl1 (*Nl-term calculation*) | 123.41 | 27.06 |
| | w3pro3md_w3xyp3 (*space propagation scheme*) | 87.01 | 19.08 |
| | w3uqckmd_w3qck3 (*time evolution scheme-3*) | 68.58 | 15.04 |
| | w3iogomd_w3outg (*output of results*) | 37.73 | 8.27 |
| | w3src2md_w3sin2 (*In-term calculation*) | 21.99 | 4.82 |
| | w3uqckmd_w3qck1 (*time evolution scheme-1*) | 17.66 | 3.87 |
| | w3srcemd_w3srce (*integration subroutine*) | 13.29 | 2.91 |
| | w3src2md_w3sds2 (*Dis-term calculation*) | 2.75 | 0.60 |
| | others | … | … |
| | **All procedures** | **455.9** | **100** |
| Modified WW | w3pro3md_w3xyp3 | 89.72 | 22.52 |
| | w3uqckmd_w3qck3 | 71.29 | 17.88 |
| | w3snl1md_w3snl1 (*Nl-term*) | 70.97 | 17.80 |
| | w3iogomd_w3outg | 38.60 | 9.68 |
| | w3uqckmd_w3qck1 | 17.97 | 4.51 |
| | w3srcemd_w3srce | 12.15 | 3.05 |
| | w3src2md_w3sds2 *(Dis-term)* | 7.68 | 1.93 |
| | w3src2md_w3sin2 *(In-term)* | 6.04 | 1.52 |
| | others | ... | ... |
| | **All procedures** | **398.8** | **100** |

From this table one can see that in the model NEW, the nonlinear term is calculated in 1.73 times faster than in the original WW. It leads to the consuming-time winning of the order of 60 seconds, which result in 15%-winning of the total consuming time. The acceleration effect is provided by using the fast DIA approximation, mentioned above. Additional small 3%-winning of time is gained due to new parameterization of the input term. But, in turn, the new approximation of *Dis*-term results in a lost of calculation speed in 2%. Nevertheless, as we said above, that just this parameterization provides, in main, the better accuracy of the model NEW, because the physics of *NL*-term and *In*-term in both models is very similar.

## 6. CONCLUSIONS

Thus, the new source function was tested and validated by means of incorporating the former into the mathematical shell of the reference model WW. Results of the test #1 are typical for any modern numerical model. But, the test #2 testifies specific properties of the proposed dissipation term. The real performance of new model was checked during the comparative validation process, which was executed in three steps differing both by duration of simulations and by regions of the World Ocean.

In general, we may state that the both models have rather high performance, which are apparently the best among present models, taking into account the results of WW's validation, presented in Tolman et al. (2002). Herewith, the comparative validation has shown a real advantage of the model NEW with respect to the original WW, especially in the accuracy of wave heights calculation. The advantage consists in reduction of the simulation errors for significant wave height, $H_s$, in 1.1-1.5 times and increasing the speed of calculation in 15%.

Analysis of the curves, like presented in Fig. 6, shows that the greatest percentage into the r.m.s. error is created at the time-series points with extreme values of wave heights and at the points corresponding to the phases while the wave intensity is going down. Both of these features are controlled by the dissipation mechanism of wave evolution. On these grounds, we conclude that the dissipation term is parameterized more efficiently in the new model than in the original WW. This property of the model NEW is very important in a sense of using it for the tasks of risk assessment.

In our study, the relative r.m.s. error, $\rho H_s$, is introduced, as one of the most instructive measure for estimation an accuracy of the wave heights simulations. In our calculations, this parameter has mean values of the order of 12-35% for both models. It is naturally to suppose that magnitudes of $\rho H_s$ should related to the value of inaccuracy of the wind field used. Regarding to this, the new task is posed, consisting in a search for a quantitative relation between errors for waves , $\rho H_s$, and the errors of input wind $\rho W_{10}$. This relation is quite expected, basing on the experimental ratios likes (33), (34). The proper study is planned to be done in a future work.

There are several another tasks related to the further validation and elaboration of wind wave numerical models. One of them consists in seeking for a certain upper limits of inaccuracy for wind field and for wave observations, which are requested for a further progress in the wind wave modeling. Estimation of these limits is the primary future task.

At present it seems that the main requirement, which define the limits of the further elaboration of numerical wind wave models, consists in using the wind field having inaccuracy below the limits mentioned.

## ACKNOWLEDGEMENTS

The work was started at the National University of Singapore in 2004 with S.A. Sannasiraj (IIT,India), continued in Russia, and finished in 2007 at CPTEC/INPE (Brasil), while Prof. V. Polnikov was standing there as the visiting scientist under support of the FAPESP, projects # 2006/56101-6. The author acknowledges the great role of Valdir Innocentini (CPTEC) in execution of calculations.